\documentstyle[twocolumn,aps]{revtex}

\input epsf

\newcommand{\dd}{\mbox{d}}

\newcommand{\gsim}{\mathrel{\raise.3ex\hbox{$>$\kern-.75em\lower1ex\hbox{$\sim$}}}}
\newcommand{\lsim}{\mathrel{\raise.3ex\hbox{$<$\kern-.75em\lower1ex\hbox{$\sim$}}}}

\begin{document}

\draft

\title{
QED corrections to deep inelastic scattering with
tagged photons at HERA
}

\author{Harald Anlauf
}
\address{
Fachbereich Physik, Siegen University, 57068 Siegen, Germany
}
\author{Andrej B. Arbuzov, Eduard A. Kuraev}
\address{
Bogoliubov Laboratory of Theoretical Physics, JINR,
141980 Dubna, Russia
}
\author{Nikolaj P. Merenkov}
\address{
Kharkov Institute of Physics and Technology,
310108 Kharkov, Ukraine
}

\date{\today}


\maketitle




\begin{abstract}
We calculate the QED corrections to deep inelastic scattering with
tagged photons at HERA in the leading logarithmic approximation.  Due to
the special experimental setup, two large scales appear in the
calculation that lead to two large logarithms of comparable size.  The
relation of our formalism to the conventional structure function
formalism is outlined.  We present some numerical results and compare
with previous calculations.
\end{abstract}

\narrowtext

\section{Introduction}
\label{sec:Intro}

One of the major tasks of the experiments at the HERA collider is the
determination of the structure functions of the proton, $F_2(x,Q^2)$ and
$F_L(x,Q^2)$, over a broad range of the kinematical variables.  The
extension of these measurements to the range of small Bjorken $x <
10^{-4}$ and $Q^2$ of a few GeV$^2$ is of particular interest, because
it will help improve our understanding of the details of the dynamics of
quarks and gluons inside the nucleon \cite{Zeuthen92}.

In order to separate $F_L(x,Q^2)$ from $F_2(x,Q^2)$, it is necessary to
measure the cross section for $ep \to eX$ with different center-of-mass
energies.  However, instead of running the collider at reduced beam
energies, one can employ a method suggested by Krasny et
al.~\cite{KPS92} that utilizes radiative events.  This method takes
advantage of a photon detector (PD) in the very forward direction, as
seen from the incoming electron beam.  Such a device is part of the
luminosity monitors of the H1 and ZEUS experiments.

The idea of the method is that emission of photons in a direction close
to the incoming electron can be interpreted as a reduction of the
effective beam energy.  The effective beam energy for each radiative
event is determined from the energy of the hard photon that is observed
in the PD.  In fact, radiative events were already used in a measurement
of the structure function $F_2$ down to $Q^2 \gsim$ 1.5~GeV$^2$
\cite{H1-96}.

The possibility to use radiative events for structure function analyses
was already discussed by Jadach {\it et al.}~\cite{JJP90}.  However, these
authors used radiative events with tagged photons in order to reduce the
radiative corrections and to simplify the $F_2$ analysis.  They did not
consider higher order corrections explicitly, and it is not clear
whether their method allows an extraction of $F_L$ without running at
lower beam energies.  On the other hand, the feasibility of a
quantitative measurement of $F_L$ at HERA using the method
of~\cite{KPS92}, for $Q^2$ below 5~GeV$^2$ and $x$ around $10^{-4}$, has
been studied in \cite{FGMZ96} and considered possible.

A more general treatment of the Born cross section for $ep \to e\gamma
X$ that is also valid for non-collinear photon emission and in the range
of high $Q^2$ can be found in the paper by Bardin {\it et al.}~\cite{BKR96}.

For an accurate description of the corresponding cross section one has
to consider the radiative corrections.  In this letter we calculate the
QED radiative corrections to deep inelastic scattering with one tagged
photon to leading logarithmic accuracy using the structure function
formalism \cite{StrFun,BBvN89}.  We will in particular discuss the
appearance of two large logarithms corresponding to two large scales,
which are due to the experimental setup peculiar to the HERA
experiments.  Finally, we will show some numerical results and compare
our findings with the calculation by Bardin {\it et al.}~\cite{BKR96}.


\section{Kinematics and lowest order cross section}
\label{sec:Kinematics}

To begin with, let us start with a brief review of the kinematics
adapted to the case of deep inelastic scattering with one exclusive hard
photon,
\begin{equation}
  e(p_e) + p(P) \to e(p_e') + \gamma(k) + X(P'),
\end{equation}
where the polar angle $\vartheta_\gamma$ of the photon (measured with
respect to the incident electron beam) is assumed to be very small,
$\vartheta_\gamma \leq \vartheta_0$, with $\vartheta_0$ being about 0.45~mrad in
the case of H1.

A convenient set of invariants that takes into account the energy loss
from the collinearly radiated photon is given by \cite{KPS92}:
\begin{eqnarray}
\label{eq:kin-vars}
        Q^2 & = & -(p_e-p_e'-k)^2 , \nonumber \\
        x   & = & \frac{Q^2}{2P \cdot (p_e-p_e'-k)} , \nonumber \\
        y   & = & \frac{P \cdot (p_e-p_e'-k)}{P \cdot (p_e-k)} .
\end{eqnarray}
Since we restrict ourselves to collinear photons, it is suitable to
parameterize the energy of the radiated photon, $E_\gamma$, with the
help of
\begin{equation}
\label{eq:z}
        z = \frac{E_e - E_\gamma}{E_e} = \frac{Q^2}{xyS},
        \qquad \mbox{ with } \quad
        S = 2 p_e \cdot P .
\end{equation}
The differential cross section for the process $ep \to e\gamma X$,
integrated over the photon emission angle $0 \leq \vartheta_\gamma \leq
\vartheta_0$, reads \cite{KPS92}
\begin{eqnarray}
\label{eq:cross-Born}
        \frac{\dd^3\sigma_{\mathrm{Born}}}{\dd x \dd Q^2 \dd z} & = &
	\sigma_0(x,Q^2;z) \cdot  \frac{\alpha}{2\pi} P(z,L_0) , \\
\noalign{\hbox{where}}
\label{eq:sigma_0}
        \sigma_0(x,Q^2;z) & = & \frac{2\pi\alpha^2}{xQ^4} \left[
        2(1-y) + \frac{y^2}{1+R} \right] F_2(x,Q^2) , \\
\noalign{\hbox{with}}
\label{eq:y}
	y & = & Q^2/(xzS) , \\
\noalign{\hbox{and}}
        P(z,L_0) & = & \frac{1+z^2}{1-z} L_0 -
        \frac{2z}{1-z}, \\
\noalign{\hbox{with}}
        L_0 & \equiv & \ln \zeta_0, \quad
        \zeta_0 = \frac{E_e^2 \vartheta_0^2}{m_e^2} .
\end{eqnarray}
Here we have neglected terms of order ${\cal O}(\vartheta_0^2)$ as well as
terms of order ${\cal O}(\zeta_0^{-1})$, and we have neglected the
contributions from $Z^0$ exchange since we are mainly interested in the
region of small $Q^2$
(see also \cite{FGMZ96}).  $R$ is defined as the ratio of cross sections
for longitudinal to transverse photons,
\begin{equation}
        R(x,Q^2) = \frac{\sigma_L}{\sigma_T} = \frac{F_L}{F_2-F_L} .
\end{equation}
The allowed range for $z$ follows from (\ref{eq:z}) and the restriction
$0 \leq y \leq 1$,
\begin{equation}
\label{eq:z-range}
        \frac{Q^2}{xS} \leq z \leq 1 .
\end{equation}
Note that the entire $z$-dependence of $\sigma_0$ occurs only via
(\ref{eq:y}).

Eq.~(\ref{eq:cross-Born}) agrees with the Born cross section given in
\cite{BKR96} when restricted to the kinematical situation under
consideration.%
\footnote{We used the same set of kinematical variables as in
\cite{KPS92} and \cite{FGMZ96}.  For details on the relation between
these two sets we refer the reader to the discussion in \cite{KPS92}.}


\section{Radiative corrections}

There are two large sources of contributions to the radiative
corrections from higher orders that contribute terms of the order of
$\alpha/\pi \cdot \ln Q^2/m_f^2$, with $m_f$ being the mass of a light
fermion.  One type of contributions are the corrections to the
propagator of the exchanged boson between the electron and the hadronic
system, while the others are due to radiation of photons off the
incoming electron line, which we will treat in the structure function
formalism.

We shall assume a calorimetric experimental setup, where a hard photon
radiated collinearly to the outgoing electron line cannot be
distinguished from a bare outgoing electron, so that final state
radiation can be neglected to the desired leading logarithmic
accuracy in accordance with the Kinoshita-Lee-Nauenberg theorem~\cite{KLN}.
We also assume some
minimum experimental cut on the transverse momentum of the outgoing
hadronic system, in order to suppress the contribution from QED Compton
events \cite{BLS93} for a leptonic measurement of the kinematic
variables.


\subsection{Vacuum polarization}
\label{sec:VP}

The first important type of contribution to the radiative corrections
comes from the vacuum polarization, which amounts to replacing the
coupling constant $\alpha$ in the hard cross section $\sigma_0$
(eq.~\ref{eq:sigma_0}) by the QED running coupling $\alpha(-Q^2)$.  For
the contribution from lepton loops we use the well-known one-loop
perturbative result.  Since we are interested in the region of rather
low $Q^2$, we use a parameterization by Burkhardt and Pietrzyk
\cite{BP95} for the hadronic contribution.


\subsection{Photonic corrections}
\label{sec:ll-corrections}

In order to illustrate the calculation of radiative corrections to the
cross section (\ref{eq:cross-Born}) due to photon emission in the
framework of the structure function formalism \cite{StrFun,BBvN89}, let
us first note that this cross section is already a QED correction: it is
that part of the inclusive first order correction to deep inelastic
scattering ($ep \to eX$), which is selected by requiring an exclusive
hard photon seen in the PD.  This may be exhibited by writing down the
radiatively corrected inclusive cross section to DIS as a convolution of
the electron non-singlet structure function $D_{\mathrm{NS}}(z,Q^2)$ with
the hard cross section (\ref{eq:sigma_0})
\begin{equation}
\label{eq:sigma-tot-inclusive}
        \frac{\dd^2 \sigma_{\mathrm{RC}}}{\dd x \dd Q^2} =
        \int \dd z \; D_{\mathrm{NS}}(z,Q^2) \, \sigma_0(x,Q^2;z) .
\end{equation}
The electron non-singlet structure function,
\begin{eqnarray}
D_{\mathrm{NS}}(z,Q^2) & = & \delta(1-z) +
        \frac{\alpha}{2\pi} \, L \, P^{(1)}(z) \nonumber \\ \label{eq:DNS}
        & + &
        \left( \frac{\alpha}{2\pi} \right)^2 \frac{L^2}{2!} P^{(2)}(z) +
        {\cal O}\left((\alpha L)^3\right),
\end{eqnarray}
depends on the large scale $Q^2$ only via the large logarithm $L = \ln
Q^2/m_e^2$.  It is known to properly sum the leading contributions
$(\alpha L)^n$ to all orders in perturbation theory \cite{BBvN89}.

We now give the relevant coefficients of the power series expansion of
$D_{\mathrm{NS}}$.  Introducing a small auxiliary parameter $\Delta$ that
serves as an infrared (IR) regulator to separate virtual+soft and hard
photon contributions, the first two coefficients of the expansion of
$D_{\mathrm{NS}}$ are
\begin{eqnarray}
\label{eq:P1}
        P^{(1,2)}(z) & = & P^{(1,2)}_\delta \cdot \delta(1-z) +
        P^{(1,2)}_\Theta(z) \cdot \Theta(1-\Delta-z) , \nonumber \\
\noalign{\hbox{with}}
        P^{(1)}_\Theta(z) & = & \frac{1+z^2}{1-z}, \qquad
        P^{(1)}_\delta = 2 \ln \Delta + \frac{3}{2} ,
\end{eqnarray}
and similarly (see e.g.\ \cite{BBvN89}),
\begin{eqnarray}
        P^{(2)}_\Theta(z) & = &
        \int_{z/(1-\Delta)}^{1-\Delta} \frac{\dd t}{t}
        P^{(1)}_\Theta(t) P^{(1)}_\Theta\left(\frac{z}{t}\right)
        + 2 P^{(1)}_\delta P^{(1)}_\Theta(z)
        \nonumber \\ & = & 2 \biggl[
        \frac{1+z^2}{1-z} \left(2\ln(1-z) - \ln z + \frac{3}{2} \right)
        \nonumber \\ \label{eq:P2}
        & + & \frac{1+z}{2} \ln z - 1 + z \biggr].
\end{eqnarray}

Inspecting (\ref{eq:cross-Born}), it is obvious that the logarithmic
piece of $P(z)$ is contained in the first order correction to the
inclusive cross section; the difference in the logarithms ($L-L_0$) is
accounted for by the remaining phase space due to photons emitted with
angle larger than $\vartheta_0$.

Let us now turn to the contributions from the second order.  Since the
maximum emission angle $\vartheta_0$ is about 0.45~mrad for the HERA
detectors, the ``large logarithm'' $L_0$ appearing in
(\ref{eq:cross-Born}) turns out to be moderate,
\[
\zeta_0 \approx 600, \quad
L_0 \approx 6.4 \qquad \mbox{at $E_e$ = 27.5~GeV} ,
\]
so that the complement
\begin{eqnarray*}
&& L_1 \equiv L - L_0 \approx 6.5 \ldots 15.7 \\
&& \mbox{(for e.g., $Q^2$ = 0.1 \ldots 1000~GeV$^2$)}
\end{eqnarray*}
is of similar magnitude or even larger than $L_0$.  For this reason we
have two large logarithms $L_0,L_1$ entering the game.

We shall separate the contributions into virtual+collinear,
soft+collinear, two collinear photons, and one collinear plus one
semicollinear photon.

The sum of the contributions of virtual+soft correction to collinear
photon emission, with the emission angle of the soft photon being
integrated over the full solid angle, but the hard photon only over the
angular range of the PD, can be obtained from the expression for the
one-loop Compton tensor.  It reads \cite{KMF87}
\begin{eqnarray}
        && \left(\frac{\alpha}{2\pi}\right)^2 L_0
	\left[ (L_0 + L_1) P^{(1)}_\delta -  (L_0 + 2 L_1) \ln z \right]
        \nonumber \\ \label{eq:V+S}
        && \qquad \times P^{(1)}_\Theta(z) \cdot \sigma_0(x,Q^2;z) .
\end{eqnarray}

The contribution from two hard photons in the PD, with the energy
fraction of each being larger than $\Delta$, is found to be \cite{Mer88}
%
\begin{eqnarray}
        && \left(\frac{\alpha}{2\pi}\right)^2 \frac{L_0^2}{2}
	\left[ P^{(2)}_\Theta(z) - 2 P^{(1)}_\Theta(z)
		\left( P^{(1)}_\delta - \ln z \right) \right]
        \nonumber \\ \label{eq:cc}
        && \qquad \times \sigma_0(x,Q^2;z) ,
\end{eqnarray}
with $z = 1 - (\sum E_\gamma)/E_e$, where $\sum E_\gamma$ is the total
photon energy registered in the PD.

Finally, we consider the contribution from one collinear photon that
hits the PD, while the other one is emitted at an angle larger than
$\vartheta_0$.

Let us investigate the regions of phase space where the large
logarithmic contributions originate.  For the hard photon that hits the
PD, the major contribution comes from the region of polar angles
$\vartheta_\gamma^{(1)} \approx m_e/E_e \ll \vartheta_0$.  Similarly, for the
other photon the biggest contribution comes from polar angles close to
the lower limit, $\vartheta_\gamma^{(2)} \gsim \vartheta_0$.  Thus, the
leading contributions come essentially from the region where
$(\vartheta_\gamma^{(2)}/\vartheta_\gamma^{(1)})^2 \approx \zeta_0 \gg 1$.
One can easily see that one obtains the double logarithms entirely from
the contribution where the photon hitting the PD is emitted first, with
the ``lost'' photon being emitted second, while the reversed case is
suppressed if the following condition is satisfied:
\begin{equation}
        x_2 \cdot \zeta_0 \gg 1,
        \qquad \mbox{ where } x_2 = \frac{E_\gamma^{(2)}}{E_e} .
\end{equation}
Thus we shall assume in the following $\Delta \ll 1$, but $\Delta \cdot
\zeta_0 \gg 1$.

Taking this ordering of photon emissions into account, the contribution
from one tagged plus one undetected photon can be calculated by means of
the quasireal electron method \cite{BFK73}
\begin{eqnarray}
        && \left(\frac{\alpha}{2\pi}\right)^2 L_0 L_1 \cdot
        P^{(1)}_\Theta(z)
        \int^{x_2^{\mathrm{max}}}_{\Delta}
        \frac{\dd x_2}{z}
        P^{(1)}_\Theta\left(1-\frac{x_2}{z}\right)
        \nonumber \\ \label{eq:sc}
        && \qquad \times \tilde{\sigma}_0(x,Q^2;z-x_2) ,
\end{eqnarray}
where $\tilde{\sigma}_0$ is understood to be expressed by the kinematic
variables $\hat x, \hat Q^2$, of the hard subprocess,
\begin{equation}
\label{eq:J}
	\tilde{\sigma}_0(x,Q^2;z-x_2) \equiv
	\sigma_0(\hat{x}, \hat{Q^2}, z-x_2) \cdot J(x,Q^2;x_2) .
\end{equation}
Note that this contribution explicitly depends on the experimental
determination of the kinematical variables $x$ and $Q^2$, since the
almost collinear emission of the second photon shifts the ``true''
kinematical variables ($\hat x, \hat Q^2$) with respect to the measured
ones ($x,Q^2$).%
\footnote{For a recent review on the influence of initial state
radiation on the experimental determination of the kinematical variables
see e.g., \cite{Wol97}, and references cited therein.}
The Jacobian $J$ accounts for scaling properties of the chosen
kinematical variables under radiation of the second photon.
The upper limit of the $x_2$-integration in (\ref{eq:sc}) is given by
either some experimental cut on the maximal energy of the second photon,
or by the kinematical limit, which also depends on the choice of the
experimental determination of the kinematical variables, as we will
discuss later.

After change of variables $x_2 = zu$, $u_0 = x_2^{\mathrm{max}}/z$, the
integral in (\ref{eq:sc}) may be conveniently decomposed into IR
divergent ($\Delta$ dependent) and IR convergent contributions as
(suppressing the arguments $x,Q^2$)
\begin{eqnarray}
\lefteqn{
        \int^{x_2^{\mathrm{max}}}_{\Delta}
        \frac{\dd x_2}{z}
        P^{(1)}_\Theta\left(1-\frac{x_2}{z}\right)
        \tilde{\sigma}_0(z-x_2)  =  } \nonumber \\
        & = &
        \int^{u_0}_{\Delta/z} \dd u \;
        P^{(1)}_\Theta(1-u)
        \left[ \left(\tilde{\sigma}_0(z(1-u)) - \tilde{\sigma}_0(z) \right)
        + \tilde{\sigma}_0(z) \right]
        \nonumber \\
        & = &
        \sigma_0(z) \cdot \left[
        \int_0^{u_0} \dd u \;
        P^{(1)}_\Theta(1-u)
        \left( \frac{\tilde{\sigma}_0(z(1-u))}{\tilde{\sigma}_0(z)} - 1 \right)
        \right. \nonumber \\ & & \left.
        \qquad \qquad
	+ 2 \ln z
        - P^{(1)}_\delta
        - \int_{u_0}^1 \dd u \; P^{(1)}_\Theta(1-u) \right],
\end{eqnarray}
where in the last step we have extended the $u$-integration of the IR
convergent piece to 0 due to the smallness of $\Delta$, and we have used
the property
\[
        \int_0^1 \dd u \; P^{(1)}(u) = 0
\]
in the simplification of the IR divergent piece.

If we sum the contributions (\ref{eq:V+S}), (\ref{eq:cc}), and
(\ref{eq:sc}), we see that the dependence on the auxiliary parameter
$\Delta$ cancels, as it should.


\subsection{The radiatively corrected cross section}

We may now write down our final result for the radiative corrections.
Since we restricted ourselves to the leading logarithmic corrections, it
follows that the radiatively corrected cross section may be written in
factorized form,
\begin{equation}
        \frac{\dd^3\sigma_{\mathrm{RC}}  }{\dd x \dd Q^2 \dd z} =
        \frac{\dd^3\sigma_{\mathrm{Born}}}{\dd x \dd Q^2 \dd z} \cdot
        \left( 1 + \delta_{\mathrm{ho}}(x,Q^2,z) \right) .
\end{equation}
where we retain in the correction factor $(1+\delta_{\mathrm{ho}})$ only
the logarithmic terms $L_0,L_1$ from (\ref{eq:V+S}), (\ref{eq:cc}), and
(\ref{eq:sc}), and the contribution from vacuum polarization,
\begin{eqnarray}
\label{eq:delta-ho}
        1 &+& \delta_{\mathrm{ho}} =
        \left(\frac{\alpha(-Q^2)}{\alpha(0)}\right)^2
        \biggl\{ 1 + \frac{\alpha L_0}{4\pi}
        \frac{1-z}{1+z^2} P^{(2)}_\Theta(z)
        \nonumber \\
        &+& \frac{\alpha L_1}{2\pi} \biggl[ \int_0^{u_0} \dd u \;
        P^{(1)}_\Theta(1-u)
        \left( \frac{\tilde{\sigma}_0(z(1-u))}{\tilde{\sigma}_0(z)} - 1 \right)
        \nonumber \\
        &-& \int_{u_0}^1 \dd u \; P^{(1)}(1-u) \biggr] \biggr\} .
\end{eqnarray}
Note that the contribution from the undetected hard photon depends on
the choice of kinematical variables and on the upper limit $u_0$.


\section{Results for HERA}
\label{sec:Results}

In the presence of the an undetected (lost) photon, the determination of
the kinematical variables $x,Q^2$ becomes ambiguous, it will depend on
the method chosen, and in turn the corrections (\ref{eq:delta-ho}) will
depend on this choice.

For HERA, several methods are being used to determine the ``true''
kinematical variables $\hat{x},\hat{Q}^2$, in order to reduce systematic
errors or to control the influence of initial state radiation (see e.g.,
\cite{Wol97} for an illuminating discussion of the latter).  For the
sake of brevity we will discuss only the so-called electron method (E)
and the Jacquet-Blondel method (JB).

In the case of the electron method, the kinematical variables
(\ref{eq:kin-vars}) are determined via
\begin{eqnarray}
        Q_e^2 & = & 4 E_e^{\mathrm{eff}} E_e' \cos^2 \vartheta_e/2,
        \nonumber \\
        y_e   & = & 1 - \frac{E_e'}{E_e^{\mathrm{eff}}} \sin^2 \vartheta_e/2,
        \quad
        x_e = \frac{Q_e^2}{y_e s^{\mathrm{eff}}},
        \\
\noalign{\hbox{with}}
        s^{\mathrm{eff}} & = & 4 E_e^{\mathrm{eff}} E_p ,
        \qquad
        E_e^{\mathrm{eff}} = E_e - E_\gamma = z E_e,
        \nonumber
\end{eqnarray}
and $E_\gamma$ is the energy deposited in the PD.  Radiating an
additional almost collinear photon with energy $E_\gamma^{(2)} = x_2
E_e$ leads to a shift of the ``true'' variables $\hat{x},\hat{Q}^2$, of
the hard subprocess with respect to the measured ones, $x_e,Q_e^2$,
\begin{equation}
\label{eq:e-method-radiative}
        \hat{Q}^2 = Q_e^2 \cdot z', \quad
        \hat{y}   = \frac{y_e + z' - 1}{z'}, \quad
        \hat{x}   = \frac{x_e y_e z'}{y_e + z' - 1},
\end{equation}
with $z' = 1 - x_2/z = 1-u$.  The kinematical limit for the undetected
photon follows from $\hat{x} \leq 1$,
\begin{equation}
        z' \geq \frac{1-y_e}{1-x_e y_e} .
\end{equation}
The Jacobian (\ref{eq:J}) for this choice of kinematical variables is
\begin{equation}
	J(x_e,Q_e^2) = \left( \frac{y_e z'}{y_e + z' - 1} \right)^2 .
\end{equation}

In the case of the Jacquet-Blondel method, the kinematical variables are
determined using
\begin{eqnarray}
        y_{\mathrm{JB}}   & = &
                \frac{1}{2E_e^{\mathrm{eff}}} \sum_h (E_h - p_{z,h}),
        \nonumber \\
        Q_{\mathrm{JB}}^2 & = &
                \frac{\sum_h p^{\perp 2}_h}{1-y_{\mathrm{JB}}},
        \quad
        x_{\mathrm{JB}}    =
                \frac{Q_{\mathrm{JB}}^2}{y_{\mathrm{JB}} s^{\mathrm{eff}}},
\end{eqnarray}
where $E_h, p_{z,h}, p_h^\perp$ are the energy, longitudinal and
transverse momentum components of the final state hadrons.

The relation between the measured and ``true'' variables in presence of
an undetected photon are now given by
\begin{equation}
\label{eq:JB-method-radiative}
        \hat{Q}^2 =
        Q_{\mathrm{JB}}^2 \frac{1-y_{\mathrm{JB}}}{1-y_{\mathrm{JB}}/z'}, \quad
        \hat{y}   = \frac{y_{\mathrm{JB}}}{z'}, \quad
        \hat{x}   =
        x_{\mathrm{JB}} \frac{1-y_{\mathrm{JB}}}{1-y_{\mathrm{JB}}/z'},
\end{equation}
with the Jacobian
\begin{equation}
        J(x_{\mathrm{JB}},Q_{\mathrm{JB}}^2) =
        \left( \frac{1 - y_{\mathrm{JB}}}{1 - y_{\mathrm{JB}}/ z'} \right)^2,
\end{equation}
and with the kinematical limit being
\begin{equation}
        z' \geq \frac{y_{\mathrm{JB}}}{1-x_{\mathrm{JB}}(1- y_{\mathrm{JB}})} .
\end{equation}

Comparing (\ref{eq:e-method-radiative}) and
(\ref{eq:JB-method-radiative}), one finds that one can in principle
determine the energy of the undetected photon (assuming that it is
emitted almost collinearly) via
\begin{equation}
\label{eq:zprime}
        z' = 1 + y_{\mathrm{JB}} - y_e .
\end{equation}
This may be used to impose an experimental cut to reduce the size of the
radiative corrections.

Let us now present some numerical results for the radiative corrections
(\ref{eq:delta-ho}).  As input we used
\begin{equation}
 E_e = 27.5 \mbox{ GeV}, \quad
 E_p = 820 \mbox{ GeV}, \quad
 \vartheta_0 = 0.45 \mbox{ mrad},
\end{equation}
the GRV94-LO structure functions \cite{GRV94} from {\tt PDFLIB}
\cite{PDFLIB} for $x > 10^{-3}$, the BK parameterization%
\footnote{We have verified that the results for the corrections do not
change significantly when we use the ALLM \cite{ALLM91} parameterization
for $x \lsim 10^{-3}$, although the difference in the predictions for
the low-$x$, low-$Q^2$ range affects the corrections already for $x$
around $10^{-4}$.}
\cite{BK92} for GRV94 for $x < 10^{-3}$, and for simplicity a fixed
value $R = 0.3$ (see e.g., \cite{H1-96} for a table of measured $R$
values).  Figure~1 shows the corrections $\delta_{\mathrm{ho}}$ for an
energy of 5~GeV deposited in the PD.

\begin{figure}[htb]
\begin{center}
\begin{picture}(80,80)
\put(5,0){
\epsfxsize=8cm
\epsfbox{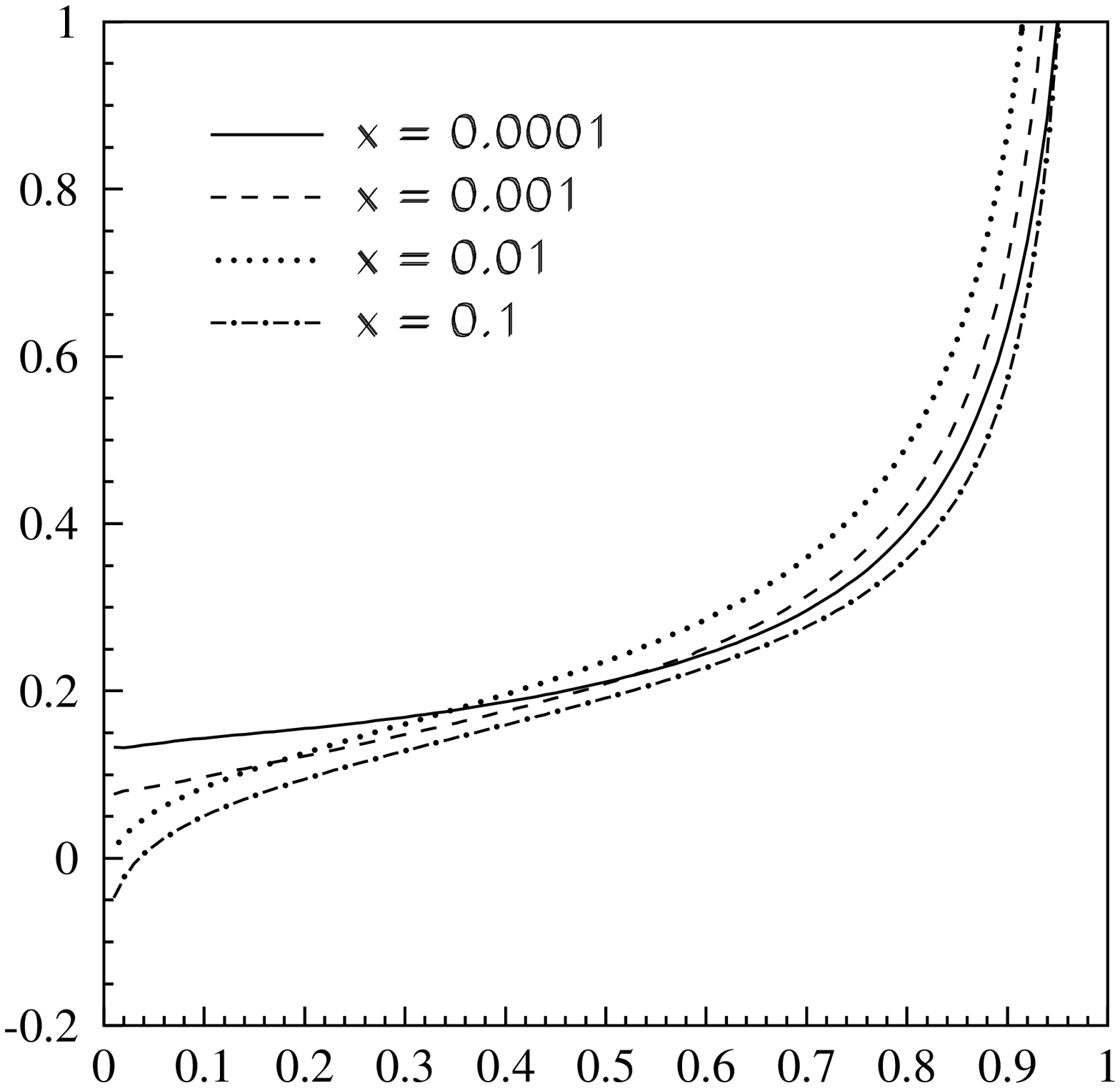}
}
\put(74,1){$y_{e}$}
\put(3,65){$\delta_{\mathrm{ho}}$}
\end{picture}
\end{center}
\caption{Radiative corrections $\delta_{\mathrm{ho}}$
(\protect\ref{eq:delta-ho}) in leptonic variables for different values
of $x$ and a tagged photon energy of 5~GeV.}
\end{figure}

At small $y_e$, the corrections are negative, since the contribution
from virtual+soft corrections dominates, because the kinematical limit
on the energy of the undetected photon tends to zero as $y_e \to 0$,
\begin{equation}
  u_0 = 1 - z'_{\mathrm{min}} =  \frac{y_e(1-x_e)}{1-x_e y_e} .
\end{equation}
For large $y_e$, the phase space for photon emission increases,
increasing also the shifts between ``true'' and measured variables
(\ref{eq:e-method-radiative}), which leads to large positive corrections
as $y_e \to 1$.

\begin{figure}[htb]
\begin{center}
\begin{picture}(80,80)
\put(5,0){
\epsfxsize=8cm
\epsfbox{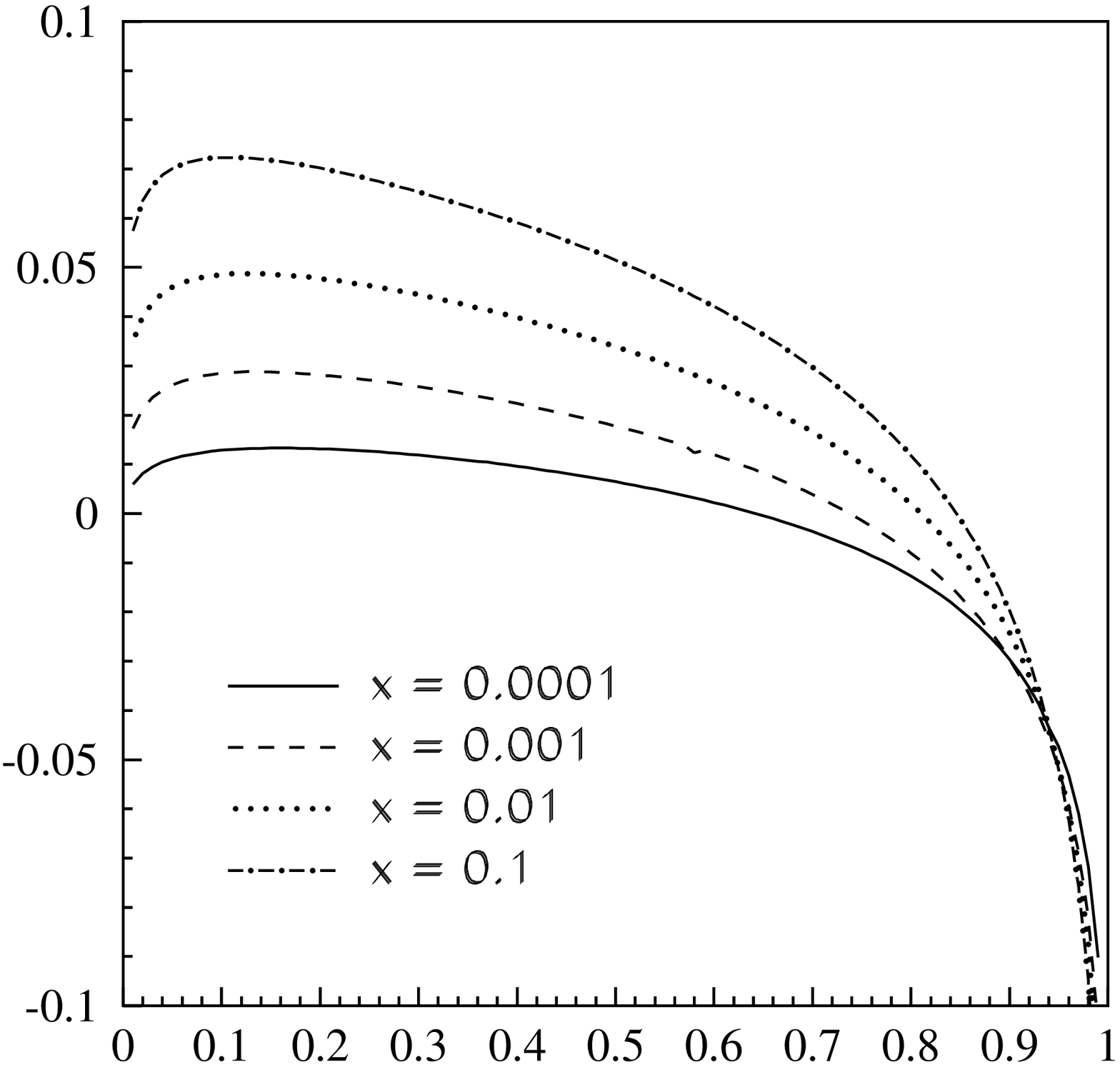}
}
\put(74,1){$y_{\mathrm{JB}}$}
\put(3,65){$\delta_{\mathrm{ho}}$}
\end{picture}
\end{center}
\caption{Radiative corrections $\delta_{\mathrm{ho}}$
(\protect\ref{eq:delta-ho}) in Jacquet-Blondel variables for different
values of $x$ and a tagged photon energy of 5~GeV.}
\end{figure}

Figure~2 shows the corrections $\delta_{\mathrm{ho}}$ for Jacquet-Blondel
variables, again for a tagged photon energy of 5~GeV.  In this case the
corrections are negative for $y_{\mathrm{JB}} \to 1$, because in this limit the
phase space for the undetected photon tends to zero,
\begin{equation}
  u_0 = 1 - z'_{\mathrm{min}} =
  \frac{(1-x_{\mathrm{JB}})(1-y_{\mathrm{JB}})}{1-x_{\mathrm{JB}}
(1-y_{\mathrm{JB}})} ,
\end{equation}
whereas for small $y_{\mathrm{JB}}$ the corrections remain moderate.
Since the Jacquet-Blondel variables correspond to a ``more inclusive''
measurement than the leptonic variables, the corrections due to
radiation of an additional photon are generally smaller.


\section{Discussion and Conclusions}
\label{sec:Conclusions}

In this letter, we have studied the radiative corrections to deep
inelastic scattering with tagged photons at HERA taking into account
only the leading logarithms.  The relevant expansion parameter in the
present case is not simply $\alpha/\pi \cdot L$, but we have two large
logarithms $L_0, L_1=L-L_0$ which happen to be of similar order of
magnitude due to the particular choice of the geometrical acceptance of
the photon detector.  The corrections do depend on the choice of the
experimental determination of the kinematical variables, but they turn
out to be of the order of 20--40\% in most regions of phase space for
leptonic variables, and below 10\% for Jacquet-Blondel variables.
Large negative corrections occur in those regions of phase space that are
dominated by soft photon emission.  These large corrections are however
well under control once one takes into account the resummation of
multiple photon emission~\cite{StrFun,BBvN89}.

We have not considered the contributions from next-to-leading order,
$(\alpha/\pi)^2 \times L$, which may be sizable since the individual
logarithms $L_0, L_1$ are not very large, especially for the
experimentally interesting region of low $Q^2$.  However, the
calculation of those terms is more involved and will be presented
elsewhere~\cite{AAK97}.

We have compared our results with the results given in \cite{BKR96}.
At the Born level, we find very good numerical agreement between
(\ref{eq:cross-Born}) and the program {\tt HECTOR}~\cite{hector}
if Z-exchange is
neglected, which is a good approximation since the experimental analysis
is restricted to the region of not too large $Q^2$ \cite{FGMZ96}.
However, at the level of radiative corrections, there are major
differences.
First, we note that our derivation of the calculation of the corrections
that is based on the structure function formalism \cite{StrFun,BBvN89}
disagrees with formula (4.1) given in their paper.  In particular, the
coefficient of the leading logarithmic term derived from (4.1) equals
two times our coefficient.  The naive usage of the quasireal electron
approximation in their formula leads to a loss of interference of
amplitudes, as was claimed by the authors.  However, the interference
actually does contribute leading logarithms when both photons hit the
PD, contrary to the statement given in \cite{BKR96}.  More details on
this interference between photon emissions can be found in \cite{Mer88}.
Furthermore, it
seems that the authors omitted the statistical factor $1/2!$, while
summing the contributions (iib, iic) of the semicollinear kinematics.

Some care is needed in the comparison of numerical results since Bardin
{\it et al.} choose kinematical variables different from ours.  Taking into
account the abovementioned factor of 2 and the difference due to our
treatment of the ordering condition for the photon emission, we can
qualitatively reproduce their results as presented in figure~6 of their
paper in the range of smaller $y$ values (thus confirming the lack of
the symmetry factor).


\section*{Acknowledgments}

We are thankful to T. Ohl and E. Boos for fruitful discussions, to
D. Bardin and L. Kalinovskaya for discussions of their paper and
sending us numerical results from the program {\tt HECTOR}~\cite{hector},
to B. Bade\l{}ek for providing us with a Fortran program for their
parameterization of $F_2$ at low $x$ and $Q^2$~\cite{BK92}.
One of us (E.K.) is grateful to the Heisenberg-Landau foundation for
financial support and to the Technische Hochschule Darmstadt for
hospitality.
Three of us (A.A., E.K., N.M.) are thankful to the INTAS foundation,
grant 93--1867 (ext.) and to INFN and the Physics Department of Parma
University for hospitality.
Finally, H.A. would like to thank
Bundesministerium f\"ur Bildung, Wissenschaft,
Forschung und Technologie (BMBF), Germany for support, and
M.~Fleischer for drawing our attention
to this problem and the members of the H1 RACO group for continued
interest.



\end{document}